\documentclass[a4paper]{article}

\usepackage{17lomcon}        
\usepackage{cite}             
\usepackage{epsfig}           

\begin{document}


\title{ ON QUASIDEGENERACY OF MAJORANA NEUTRINOS AND THE OBSERVED  PATTERN OF LEPTONIC MIXING}

\author{M. N. Rebelo\email{rebelor@tecnico.ulisboa.pt}}

\affiliation{Centro de F\' \i sica Te\' orica de Part\' \i culas - CFTP and Departamento de F\' \i sica, 
Instituto Superior T\' ecnico - IST, Universidade de Lisboa, Av. Rovisco Pais, P-1049-001 Lisboa,  Portugal}


\date{}
\maketitle


\begin{abstract}
We relate the observed pattern of leptonic mixing to the quaidegeneracy of three 
Majorana neutrinos. We show how lifting the degeneracy may lead to the measured value 
of $|U_{13}|$ and to sizeable CP violation of Dirac-type. We show some of the
correlations obtained among physical observables, starting from some of the most interesting
schemes proposed in the literature.
\end{abstract}

\section{Introduction}
This talk is based on work done in collaboration with G. C. Branco, J. I. 
Silva-Marcos and Daniel Wegman \cite{Branco:2014zza}.

It was already shown, quite some time ago that the limit of exact degeneracy
of Majorana neutrinos is non trivial in the sense that it allows for leptonic mixing and
CP violation \cite{Branco:1998bw}. This result relies on the assumption that 
neutrinos are Majorana particles. Only fermions that are neutral under all
$U(1)$ type symmetries can have Majorana character \cite{Mohapatra:1998rq}, 
therefore all other fermions in the 
Standard Model (SM) are Dirac fermions. Despite great recent experimental progress
in the field of Neutrino Physics, the question of whether neutrinos are Dirac or
Majorana fermions still remains one of the major open questions in Particle
Physics.

The 2015 Nobel prize was awarded to T. Kajita and B. McDonald
``for the discovery of neutrino oscillations, which shows that neutrinos
have mass". In fact, it is by now established that at least two of the
observed neutrinos have non-zero masses. The presently allowed ranges are listed 
in Table 1.  It should be noted that the sign of  $\Delta m^2_{31} \equiv m^2_3 - m^2_1$,
is not yet known and is at present under intense experimental investigation. 
Experimentally the mass of the lightest neutrino, either $\nu_1$ or $\nu_3$ 
is still consistent with zero. However,
upper bounds for the neutrino masses allow for quasidegeneracy. 
It is therefore of great interest to consider the limit of exact degeneracy of
neutrino masses and possible ways of lifting this degeneracy. Specially so
because the limit of exact degeneracy favours two large mixing angles 
and leads to one zero mixing angle which upon the lifting of  the degeneracy can give rise
to the observed small angle. The leptonic mixing angles have  been measured 
experimentally, and are also listed in Table 1. There are two angles that are large when 
compared to the Cabibbo angle and one small mixing angle, $\theta_{13}$, which was consistent
with zero until the recent measurements performed at reactor \cite{reactor} and accelerator
neutrino experiments  \cite{Abe:2013xua}.  Leptonic mixing is remarkably different from quark 
mixing. It is not yet known whether or not there is CP violation in
the leptonic sector either at low or at high energies \cite{Branco:2011zb}.  
In the absence of a flavour model
no relation can be established between these two manifestations \cite{Branco:2001pq},
\cite{Rebelo:2002wj}. The limit of exact degeneracy with Majorana
neutrinos allows for one Majorana-type CP violating phase in the mixing
matrix  \cite{Branco:1998bw}. Lifting of the degeneracy together with the requirement
that all three mixing angles should be different from zero allows for 
Dirac as well as Majorana-type CP violation \cite{Branco:2014zza}.  In Ref. \cite{Branco:2014zza}  
we studied  perturbations of exact degeneracy
around some of the very well known mixing textures 
proposed in the literature, with $\theta_{13} = 0$,  \cite{Harrison:2002er}, 
\cite{Fritzsch:1995dj}, \cite{Barger:1998ta}, \cite{Kajiyama:2007gx}, \cite{Rodejohann:2008ir},
\cite{Albright:2010ap}, \cite{Kim:2010zub}, \cite{Kim:2011vr}
leading to quasidegenerate masses and 
mixing in agreement with all available experimental data. We also showed that in this 
process it is possible to generate  Dirac-type CP violation large enough in strength to be  
detectable in the next round of neutrino experiments
without introducing new sources of CP violation.
  
\section{Present Experimental Knowledge of Neutrino Masses and Leptonic Mixing}
Table 1 gives the 2014 update on global fits of neutrino oscillation parameters 
provided by Forero, Tortola and Valle \cite{Forero:2014bxa}.  The quantities
$\Delta m^2_{ij}$ are defined by $(m^2_i - m^2_j)$ and the angle $\theta_{ij}$
and the phase $\delta$  are those of the standard parametrisation \cite{Agashe:2014kda}
\begin{center}
\begin{table}[h]
\caption{Neutrino oscillation parameter summary. For $\Delta m^2_{31}$, 
$\sin^2 \protect\theta_{23}$ , $\sin^2 \protect\theta_{13}$, and $\protect
\delta$ the upper (lower) row corresponds to normal (inverted) neutrino mass
hierarchy.
${}^a$There is a local minimum in the first octant, $\sin^2 \theta_{23}= 0.467 $
with $\Delta \chi^2 =0.28$ with respect to the global minimum. From
\cite{Forero:2014bxa}}
\label{reps}
\begin{tabular}{ccc}
\hline\hline
Parameter & Best fit & $1 \sigma $ range \\ \hline
$\Delta m^2_{21}$ $[10^{-5} eV^2 ] $ & 7.60 & 7.42 -- 7.79 \\ 
$|\Delta m^2_{31}|$ $[10^{-3} eV^2 ] (NH)$ & 2.48 & 2.41 -- 2.53 \\ 
$|\Delta m^2_{31}|$ $[10^{-3} eV^2 ] (IH) $ & 2.38 & 2.32 -- 2.43 \\ 
$\sin^2 \theta_{12}$ & 0.323 & 0.307 -- 0.339 \\ 
$\sin^2 \theta_{23} (NH)$ & 0.567  & $0.439^a$ -- 0.599 \\ 
$\sin^2 \theta_{23} (IH)$ & 0.573 & 0.530 -- 0.598 \\ 
$\sin^2 \theta_{13}$ (NH)& 0.0234 & 0.0214 --0.0254 \\ 
$\sin^2 \theta_{13}$ (IH) & 0.0240 & 0.0221 -- 0.0259 \\ 
$\delta$ (NH) & 1.34 $\pi $ & 0.96 --1.98 $\pi$ \\ 
$\delta$ (IH) & 1.48 $\pi$ & 1.16 --1.82 $\pi$ \\ \hline
\end{tabular}
\end{table}
\end{center}
Several textures for the leptonic mixing have been studied in the literature,
often in the context of family symmetries \cite{Altarelli:2010gt}, 
\cite{King:2014nza}, \cite{Fonseca:2014koa}. 
In most of the proposed schemes, the pattern of leptonic mixing is predicted 
but the spectrum of masses is not constrained by
the symmetries. It is therefore consistent to consider these schemes, together with
the hypothesis of quasidegeneracy of Majorana neutrinos.

\section{The Limit of Exact Degeneracy with Majorana Neutrinos}
We work in the framework of three left-handed neutrinos and write the effective
Majorana mass term in the weak basis where the charged lepton mass matrix is 
diagonal, real and positive as:
\begin{equation}
\mathcal{L}_{{\mathrm{mass}}}\ =\ -\ ({\nu }_{{_{L}}_{\alpha }})^{T}\
C^{-1}\ (M_{o})_{\alpha \beta \quad }{\nu }_{{_{L}}_{\beta }}\ +\mathrm{\
h.c.}  \label{eq1}
\end{equation}
where ${\nu }_{{_{L}}_{\alpha }}$ stand for the left-handed weak eigenstates
and $M_{o}$ is a $3\times 3$ symmetric complex mass matrix.
Obviously there is no loss of generality in choosing this weak-basis.  In general
$M_{o}$ is diagonalized by a unitary matrix $U_{o}$ through $U_{o}^{T}\
M_{o}\ U_{o}$ $=diag$ $(m_{\nu _{1}},m_{\nu _{2}},m_{\nu _{3}})$. In
the limit of exact degeneracy, $M_{o}$ can be written: 
\begin{equation}
M_{o}={\mu }\ S_{o}
\end{equation}
where ${\mu }$ is the common neutrino mass and $S_{o}=U_{o}^{\ast
}U_{o}^{\dagger }$. Therefore, in the limit of exact degeneracy
 $M_{o}$ is proportional to a symmetric unitary matrix. 
Leptonic mixing and even CP violation can occur in this limit provided
that neutrinos are Majorana particles  \cite{Branco:1998bw}. 
Leptonic mixing can be rotated away if and only if there is CP invariance and all
neutrinos have the same CP parity \cite{Wolfenstein:1981rk}.
It was also shown in Ref \cite{Branco:1998bw} that in the case of different CP parities
the leptonic
mixing matrix $U_{o}$ can be parametrised by two angles and one phase, in the form: 
\begin{equation}
U_{o}\ =O_{23}\left( {\small \phi }\right) {\small \ }O_{12}\left( 
\frac{{\small \theta }}{{\small 2}}\right) \ \left( 
\begin{array}{ccc}
1 & 0 & 0 \\ 
0 & i & 0 \\ 
0 & 0 & e^{-i\frac{\alpha }{2}}
\end{array}
\right)  \label{eq8}
\end{equation}
up to an orthogonal rotation of the three degenerate neutrinos and with
each orthogonal matrix $O_{ij}$ chosen to be symmetric. Obviously, if 
$U_{o}$ diagonalises $M_0$ so does $U_{o} O$ with $O$ an 
arbitrary orthogonal rotation. An important feature is the fact that
$U_{o}$ always has one zero
entry which in the above parametrisation appears in the $(13)$ position.
Although the location of the zero is not fixed the observed pattern of leptonic
mixing suggests that the above choice is a good starting point for a perturbation. 

\section{Lifting the Degeneracy}
In order to lift the degeneracy one may add a small perturbation to $S_0$:
\begin{equation}
M={\mu }\ (S_{o}+\varepsilon ^{2}\ Q_{o})  \label{quasi}
\end{equation}
where the matrix $Q_0$ is fixed in such a way that the correct neutrino masses are
obtained.
We assume that after lifting the degeneracy the leptonic mixing matrix is given by:
\begin{equation}
U_{PMNS}=U_{o}\cdot O  \label{uuo}
\end{equation}
where $O$ is an orthogonal matrix, parametrized by small angles. The matrix 
$O$ is denoted by:
\begin{equation}
O=O_{12}O_{13}O_{23}=\left( 
\begin{array}{ccc}
c_{\phi _{1}} & s_{\phi _{1}} & 0 \\ 
-s_{\phi _{1}} & c_{\phi _{1}} & 0 \\ 
0 & 0 & 1
\end{array}
\right) \left( 
\begin{array}{ccc}
c_{\phi _{3}} & 0 & s_{\phi _{3}} \\ 
0 & 1 & 0 \\ 
-s_{\phi _{3}} & 0 & c_{\phi _{3}}
\end{array}
\right) \left( 
\begin{array}{ccc}
1 & 0 & 0 \\ 
0 & c_{\phi _{2}} & s_{\phi _{2}} \\ 
0 & -s_{\phi _{2}} & c_{\phi _{2}}
\end{array}
\right)  \label{ooo}
\end{equation}
Notice that 
$U_{PMNS}$ still diagonalises $S_{o}$, thus establishing a strong connection 
between the degenerate and quasidegenerate case. 
Different cases were analysed by choosing  $U_0$  to coincide with some of the
most interesting cases considered in the literature with a zero in the $(13)$ entry. 

One of the examples considered in Ref.~\cite{Branco:2014zza} consists of perturbing
the tribimaximal mixing. In this case, we have:
$U_{o}=U_{TBM}\cdot K$ with: 
\begin{equation}
U_{TBM}=\left( 
\begin{array}{ccc}
\frac{2}{\sqrt{6}} & \frac{1}{\sqrt{3}} & 0 \\ 
\frac{1}{\sqrt{6}} & -\frac{1}{\sqrt{3}} & \frac{1}{\sqrt{2}} \\ 
\frac{1}{\sqrt{6}} & -\frac{1}{\sqrt{3}} & -\frac{1}{\sqrt{2}}%
\end{array}%
\right) \qquad \mbox{and}\qquad K=diag(1,i,e^{-i\alpha /2})
\end{equation}%
In the notation of Eq.~(\ref{eq8}), this ansatz corresponds to $\phi
=45^{\circ }$ and $\cos \left( \frac{\theta }{2}\right) =\frac{2}{\sqrt{6}}$
i.e., $\frac{\theta }{2}=35.26^{\circ }$.  We used data from the global fit of 
neutrino oscillations provided in 2012 by the authors of Ref.~\cite{Forero:2014bxa},
requiring agreement within $1 \sigma$ range. Although there are deviations in the more 
recent data from that of 2012 the deviations are slight  and therefore the conclusions
would not change significantly. In Ref.~\cite{Branco:2014zza}
we concluded that, in this example, complying with the  experimental bounds allowed 
the leptonic strength 
of Dirac-type CP violation to range from  0 to about $4\times 10^{-2}$, so that it could be 
within reach of future neutrino experiments.

Here we reproduce one of the figures obtained in the previous reference showing the correlation 
between $I_{CP}$ and $|U_{13}|^2$. This scenario allows for a particularly simple solution 
since, one can reach agreement with the experimental data by choosing a matrix $O$ with only one
parameter different from zero, namely the angle $\phi _{2}$.
\begin{figure}[h]
\begin{center}
\includegraphics[scale=0.5]{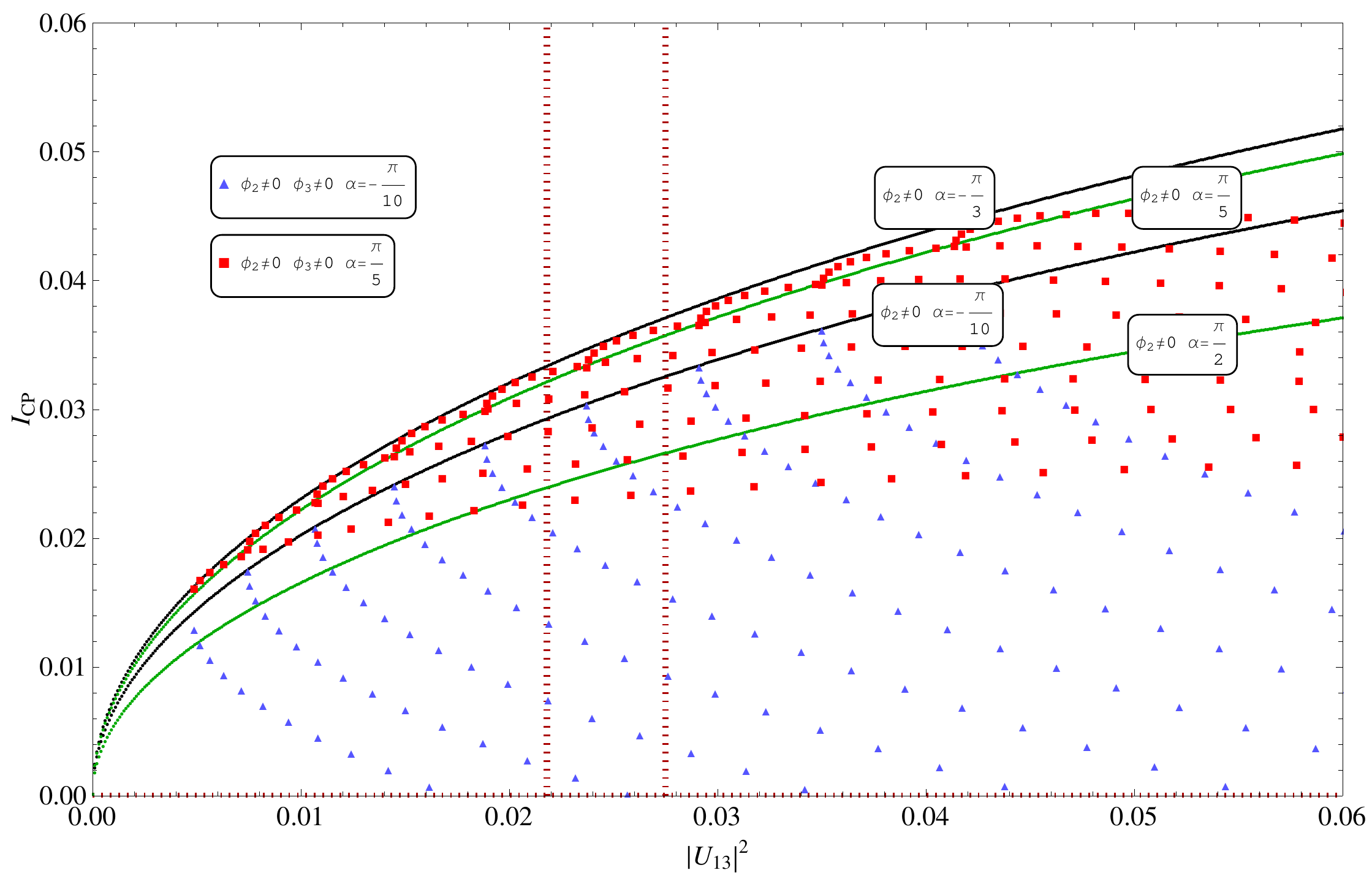}
\end{center}
\caption{$I_{CP}$ versus $|U_{13}|^{2}$ obtained by perturbing tribimaximal
mixing with $\protect\phi _{3}=0$. Each curve corresponds to a fixed $%
\protect\alpha $ and to $\protect\phi _{1}=0$ , therefore $\protect\phi _{2}$
is the only variable. The points drifting away from each curve were obtained
by varying also $\protect\phi _{3}$.}
\label{figura3}
\end{figure}

\section{Final Remarks}
The starting point of our analysis is the limit of exact degeneracy of neutrino masses.
In this limit there is no Dirac-type CP violation, only Majorana-type CP violation may occur
and one of the leptonic mixing angles is exactly zero.
It has been shown in Ref.~ \cite{Emmanuel-Costa:2015tca} that there is no loss of generality in 
parametrising  a general unitarity matrix in the form given by Eq.(\ref{uuo}), with $U_0$ 
of the generic form introduced before,
however fixing the parameters of  $U_0$, as was done in the previous analysis, by
making use of some of the most interesting patterns discussed in the literature with
$\theta_{13}=0$ restricts  $U_{PMNS}$, 
and gives rise to  predictive power while at the same time allowing to relate 
quasidegeneracy of neutrino masses with the observed pattern of leptonic mixing..

\section*{Acknowledgments}

The author thanks the Organisers of the 17th Lomonosov Conference on Elementary Particle 
Physics for the very stimulating scientific environment provided during the Conference as well as the 
warm hospitality. This work was partially supported by Funda\c c\~ ao para a Ci\^ encia e a 
Tecnologia (FCT, Portugal) through the projects PTDC/FIS-NUC/0548/2012, 
CERN/FIS-NUC/0010/2015,  CFTP-FCT Unit 777 (UID/FIS/00777/2013) which are partially 
funded through POCTI (FEDER),  COMPETE, QREN and EU. Part of the work presented here 
was done at CERN-Theory Division during different periods of time and with partial financial support from CERN.


\end{document}